\documentclass[runningheads]{llncs}
\usepackage[T1]{fontenc}
\usepackage{graphicx}
\usepackage{amsfonts}
\usepackage{amsmath}
\usepackage{cases}
\usepackage[caption=false]{subfig}
\usepackage{xcolor}
\usepackage{algorithm}
\usepackage[noend]{algpseudocode}
\makeatletter
\def\BState{\State\hskip-\ALG@thistlm}
\makeatother

\newcommand{\E}{\mathbb{E}}

\newtheorem{assumption}{Assumption}

\begin{document}
\title{A Multi-Resolution Dynamic Game Framework for Cross-Echelon Decision-Making in Cyber Warfare}
\titlerunning{Multi-Resolution Game for Cross-Echelon Decision-Making}

\author{Ya-Ting Yang  \and
Quanyan Zhu }

\institute{
New York University, Brooklyn, NY, USA\\ 
E-mail: \email{\{yy4348, qz494\}@nyu.edu}}

\maketitle              
\begin{abstract}
Cyber warfare has become a critical dimension of modern conflict, driven by society’s increasing dependence on interconnected digital and physical infrastructure. Effective cyber defense often requires decision-making at different echelons, where the tactical layer focuses on detailed actions such as techniques, tactics, and procedures, while the strategic layer addresses long-term objectives and coordinated planning. Modeling these interactions at different echelons remains challenging due to the dynamic, large-scale, and interdependent nature of cyber environments. To address this, we propose a multi-resolution dynamic game framework in which the tactical layer captures fine-grained interactions using high-resolution extensive-form game trees, while the strategic layer is modeled as a Markov game defined over lower-resolution states abstracted from those game trees. This framework supports scalable reasoning and planning across different levels of abstraction through zoom-in and zoom-out operations that adjust the granularity of the modeling based on operational needs. A case study demonstrates how the framework works and its effectiveness in improving the defender's strategic advantage.
\end{abstract}
\section{Introduction}

Cyber warfare refers to the use of cyber capabilities to disrupt, degrade, or destroy an adversary’s information systems and digital or physical infrastructure in pursuit of strategic objectives \cite{7529426}. Its primary objectives span both the public and private sectors, including government and military networks, critical infrastructure such as power grids and water treatment systems, and essential civilian services such as healthcare, finance and telecommunications \cite{acton2020cyber}. As modern societies become increasingly dependent on these interconnected systems, cyber warfare has emerged as a critical concern across academia, industry, and government.

Decision-making in the cyber domain occurs across multiple echelons, typically categorized into tactical and strategic layers \cite{li2024symbiotic}. Tactical decision-making focuses on the implementation of specific techniques, tactics, and procedures (TTPs) used in individual attacks and defenses. Established frameworks like MITRE ATT\&CK \cite{strom2018mitre} provide structured guidance at a higher resolution of detail, helping practitioners select and orchestrate appropriate actions. Strategic decision-making involves broader planning and coordination, designing sequences of tactical operations, allocating resources, and aligning cyber activities with overarching mission goals. A key example is defensive cyber deception \cite{javadpour2024comprehensive}, in which defenders manipulate attacker perception and behavior by shaping the observable environment. At the tactical layer, this can involve deploying honeypots, fake credentials, or simulated network traffic. At the strategic layer, such tactics must be coordinated to support long-term defense objectives, such as misdirecting attackers, gathering intelligence, or shaping adversary beliefs about the system.

\begin{figure}[htp]
\centering \vspace{-5mm}
\subfloat[ \label{fig:intro}]{%
  \includegraphics[width=0.4\textwidth]{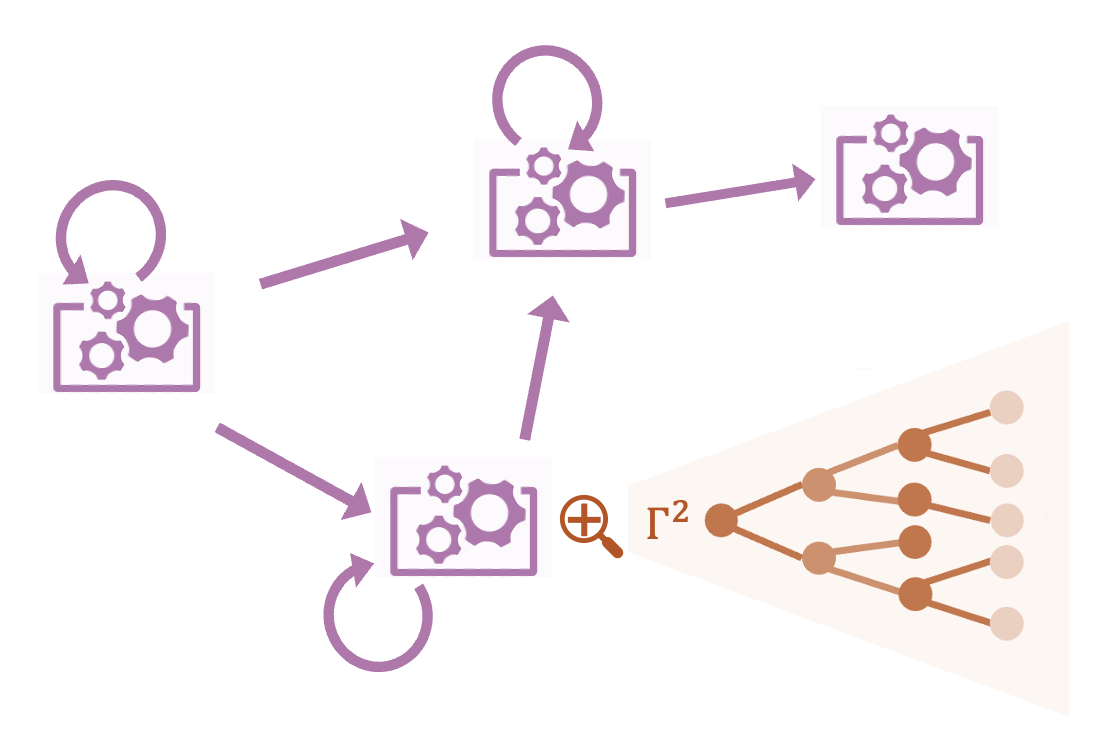}%
}\hfil
\subfloat[ \label{fig:intro_}]{%
  \includegraphics[width=0.6\textwidth]{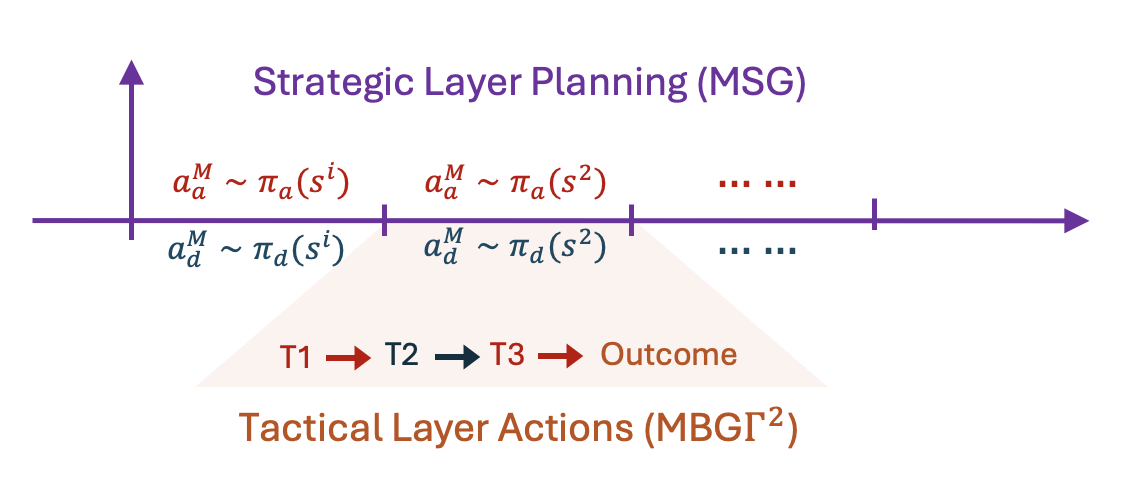}%
}
\caption{An illustration of multi-layered decision-making ($s$ for lower and $\Gamma$ for higher resolution). (a) The MSG abstracts the system considered during cyber warfare as a network of interconnected MBGs. (b) The strategic layer governs inter-MBG planning at a coarser resolution, while the tactical layer focuses on detailed action sequences within each MBG.}
\vspace{-1mm}
\label{fig:int}
\end{figure}

Given the multi-layered nature of decision-making in cyber operations, cross-echelon coordination, which integrates both tactical and strategic reasoning, is essential for achieving effective and resilient defenses. In practice, a defender may incur a tactical loss, such as allowing an attacker to breach the decoy system, but still secure a strategic advantage, such as protecting critical assets by diverting the attacker’s efforts. This asymmetry between short-term setbacks and long-term objectives highlights the inherent complexity of cyber warfare, where one decision made at one layer can influence outcomes of other decisions at the same or another layer. Effectively addressing this complexity requires a principled framework that captures varying levels of resolution and the interdependencies between detailed engagements and broader campaign goals.

Since cyber warfare involves strategic interactions between adversarial entities, such as attackers and defenders with conflicts of interest, game-theoretic frameworks \cite{kamhoua2021game} naturally serve as effective tools for modeling and analyzing their behavior. However, several fundamental key issues remain. First, cyber environments are inherently dynamic, with rapidly evolving threats and shifting attack surfaces \cite{mallick2024navigating}. Second, the complexity of cyber operations is increasing due to the scale of enterprise networks, the interconnectivity and heterogeneity of digital assets, and the wide range of potential attack tactics and objectives \cite{wen2017complex}. Planning in such settings often requires reasoning over large, multi-stage decision spaces, where outcomes depend on sequences of interdependent actions and delayed effects. These challenges are further amplified by cross-echelon interactions, making conventional static or dynamic but single-resolution models insufficient to capture the full scope of cyber warfare.

To address these challenges, we propose a multi-resolution dynamic game framework for supporting multi-echelon decision-making in cyber warfare. As illustrated in Figure \ref{fig:int}, at the tactical layer, microbase games (MBGs) are modeled using extensive-form game trees \cite{seq_eqm} to capture step-by-step high resolution interactions between players. In the strategic layer, the macro-strategic game (MSG) is formulated as a Markov game \cite{bacsar1998dynamic}, which abstracts the system into a network of interconnected MBGs. This macro-layer captures the interdependencies between local engagements and provides lower-resolution insights that remain scalable and computationally tractable as the number of interactions at the tactical layer grows. By incorporating formally defined zoom-in and zoom-out operations, the framework enables dynamic adjustment of resolution based on operational needs. This enables purple teaming and allows the defensive entity to explore specific components of the system in greater detail when necessary, while maintaining an overview to support strategic planning and cross-layer coordination. Our case study further demonstrates that the multi-resolution operations help the defensive entity gain improved strategic advantage, both within individual MBGs and across the overarching MSG.

\section{Related Work}

\subsection{Cyber Warfare}
The nature of cyber warfare distinguishes it from traditional military warfare \cite{cohen1996revolution} through its anonymity, asymmetry, and capacity to operate below the threshold of open hostilities \cite{robinson2015cyber}. Cyber attacks can be launched remotely, enabling plausible deniability and complicating attribution, which hinders timely and coordinated responses. High-profile incidents, such as Stuxnet \cite{farwell2011stuxnet}, the 2007 Estonia attacks \cite{ottis2008analysis}, and the 2015 Ukraine power grid breach \cite{case2016analysis}, demonstrate how cyber operations can produce strategic disruption across both digital and physical domains. For example, Advanced Persistent Threats (APTs) exemplify long-term, stealthy campaigns targeting espionage or critical infrastructure disruption \cite{alshamrani2019survey,ge2024mega}. In response, cyber deception has emerged as a key defensive measure, using decoys, fake credentials, and simulated environments to mislead and delay adversaries \cite{javadpour2024comprehensive,horak2017manipulating,zhang2018hypothesis,yang2025deceive}. At the same time, the rapid advancement of technologies such as artificial intelligence \cite{johnson2019artificial,hartmann2020next} and quantum computing \cite{krelina2021quantum,radanliev2025cyber} has significantly amplified the precision, scale, and autonomy of cyber operations. Despite these trends, most of the existing research remains focused on technical and tactical defense solutions, with relatively limited emphasis on integrated frameworks that can inform cross-echelon cyber decision making.

\subsection{Game-Theoretic Decision-Making}
One of the core challenges in modeling interactions through dynamic games such as extensive-form games and Markov games in cyber operations is the curse of dimensionality: as the number of decision stages, system states, and possible actions increases, the computational complexity increases rapidly, making analysis and planning intractable \cite{pakes2001stochastic,doraszelski2012avoiding}. To address this issue, prior research has explored techniques such as state aggregation \cite{ren2002state}, action abstraction \cite{marino2019evolving}, and function approximation \cite{lagoudakis2012value} to approximate large-scale Markov decision processes (MDP) or games, typically solved using methods from approximate dynamic programming (ADP) \cite{powell2007approximate}. Although effective in reducing computational complexity, these approaches often sacrifice interpretability and operational relevance, as the abstraction process obscures detailed insights from the original model. Motivated by multigrid optimization methods \cite{nash2000multigrid}, which operate across multiple levels of resolution (or ``grids'') to accelerate convergence and reduce computational costs, this work introduces a multi-resolution dynamic game framework that leverages the MSG for scalable planning, while retaining the ability to ``zoom in'' and then ``zoom out'' on MBGs when detailed tactical reasoning or plan refinement is needed. This enables both computational tractability and operational transparency across different layers of decision-making.

\section{The Micro-Level Game}
To construct the proposed multi-resolution game, we start with the high-resolution base game at the micro-level, which captures detailed (tactical) interactions between players. These micro-level base games serve as foundational components for constructing the macro level, enabling a multi-resolution framework that bridges fine-grained dynamics with higher-level strategic insights.
\subsection{The Base Game}
To model the sequential actions between two players who have a conflict of interest, such as malicious player $P_A$ and defensive player $P_D$ within the base game, we adopt the concept of an extensive-form game tree from \cite{seq_eqm}. This approach explicitly and visually represents the sequential moves, possible outcomes, and available information at each decision point during the strategic interactions between players.
Mathematically, the extensive-form game formulation is constructed from the following elements.
\paragraph{The physical order of play $(\mathcal{T}, \prec)$:} A finite set $\mathcal{T}$ of (tree) nodes for the game tree together with a binary relation $\prec$ on $\mathcal{T}$ representing precedence. One node $t \in \mathcal{T}$ precedes another if there is a unique path (formed by a sequence of actions taken) from the former to the latter. The binary relation $\prec$ must be a partial order, and $(\mathcal{T}, \prec)$ must form an arborescence\footnote{This helps prevent cycles from appearing in the order of play, and it means that each node in the tree can be reached by one and only one path from an initial node through the tree.}: the relation $\prec$ totally orders the predecessors of each member of $\mathcal{T}$. Some auxiliary notations are then defined as follows:
\begin{itemize}
    \item $\mathcal{Z}=\{t \in \mathcal{T}:\text{node $t$ has no successors}\}$ denotes the set of terminal nodes or the outcomes for the game tree.
    \item $\mathcal{X}=\mathcal{T}\setminus \mathcal{Z}$ denotes the set of decision nodes.
    \item $\mathcal{W}=\{t \in \mathcal{T}:\text{node $t$ has no predecessors}\}$ represents the set of initial nodes or states.
    \item  $p(t)=\{x \in \mathcal{X}: x \prec t\}$: predecessors of node $t$.
    \item $p_I(t)=\max \{x \in \mathcal{X}: x \prec^ t\}$ for $t \notin \mathcal{W}$: immediate predecessors of node $t$.
    \item $p_n(t)=p_I(p_{n-1}(t))$ for $t$ such that $p_{n-1}(t) \notin \mathcal{W},$ and $p_{0}(t)=t$ for all $t$: the $n$-th predecessors of node $t$.
    \item $l(t)$ such that $p_{l(t)}(t) \in \mathcal{W}$: number of predecessor of node $t$.
    \item $m(x)=(p_I)^{-1}(x)$ for $x \in \mathcal{X}$: immediate successors of decision node $x$.
    \item $z(x)=\{z \in \mathcal{Z}: x \prec z\}$ for $x \in \mathcal{X}$: terminal successors of decision node $x$.
\end{itemize}
With these notations, the base game begins at one of the initial nodes (determined by nature) and then proceeds along some path from the node to an immediate successor, terminating when a terminal node is reached.

\paragraph{Players and Turn function $(\mathcal{N}, I)$:}
A set of players $\mathcal{N}=\{P_A, P_D\}$ and a function $I:\mathcal{X} \mapsto \mathcal{N}$ that assigns to each decision node $x$ the player whose turn it is.

\paragraph{Choices available $(\mathcal{A}, \alpha)$:}
A finite set $\mathcal{A}$ of actions and a function $\alpha: \mathcal{T}\setminus \mathcal{W} \mapsto \mathcal{A}$ that labels each non-initial node with the last action taken to reach it. Here, $\alpha(m(x))$ is the set of feasible actions at the decision node $x \in \mathcal{X}$, and $\alpha$ is required to be a one-to-one on the set $m(x)$ of immediate successors of $x$.

\paragraph{Information processed $\mathcal{H}$:}
A partition $\mathcal{H}$ of $\mathcal{X}$ that divides the decision nodes into information sets. The cell $H(x)$ of $\mathcal{H}$ that contains $x$ identifies the decision nodes that the player $I(x)$ cannot distinguish from $x$ based on the information available when it is his/her turn to choose an action in $x$. It is required that a player knows when it is his/her turn and which actions are available. That is, if $x \in H(x')$, then
$$
I(x)=I(x') \text{ and } \alpha(m(x))=\alpha(m(x')).
$$ Hence, we can write $I(h)$ for $h \in \mathcal{H}$ and partition $\mathcal{H}$ into sets $\mathcal{H}_i=I^{-1}(P_i)$. That is, $\mathcal{H}_i$ is the set of information sets at
which player $P_i$ moves. More formally, 
$$
\mathcal{H}_i = \{\mathcal{M} \subset \mathcal{X}: \mathcal{M}=H(x) \text{ for } x \in \mathcal{X} \text{ with } I(x) = P_i\}.
$$ Each $h \in \mathcal{H}_i$ represents a set of decision nodes where the player $P_i$ has the same feasible actions $A(h)=\alpha(m(h))$ and cannot distinguish between the nodes within $h$. Then, we denote $A_i=\{A(h)\}_{h \in \mathcal{H}_i}$ for the set of actions available to player $P_i$ at any of his/her information sets.

To this end, the extensive form for the micro base game is defined by the collection $\Xi=\langle \mathcal{T}, \prec; \mathcal{N}, I; \mathcal{A}, \alpha; \mathcal{H} \rangle$. Using this, we can then have the definition for the base game as follows.

\begin{definition}[Micro Base Game (MBG)]
\label{def:MLG}
The Base Game at the micro level can be defined by an extensive-form game tuple $\Gamma = \langle \Xi, \sigma_c, r_A, r_D\rangle$, where each component represents:
\begin{itemize}
    \item $\Xi=\langle \mathcal{T}, \prec; \mathcal{N}, I; \mathcal{A}, \alpha; \mathcal{H} \rangle$ is the extensive form. In addition, $\mathcal{Z} \subset \mathcal{T}$ represents the finite set of possible outcomes for MBG. 
    \item $\sigma_c \in \Delta(\mathcal{W})$ is a probability measure on the set $\mathcal{W}$ of states or initial nodes, as for notational convenience, we have put all actions by nature at the ``start'' of the game. That is, $\sigma_c$ is nature's fixed policy. 
    \item $r_A: \mathcal{Z} \mapsto \mathbb{R}, r_D: \mathcal{Z} \mapsto \mathbb{R}$ are the utility functions for players $P_A$ and $P_D$, respectively, which determine the payoffs or costs the players receive when reaching a certain outcome.
\end{itemize}
\end{definition}

\subsection{Strategies and Solution Concepts}

Given the MBG, players can adopt different types of strategies depending on how they choose actions throughout the game tree. We begin by considering pure strategies.
\begin{definition}[Micro Pure Strategy]
    Consider the MBG $\Gamma$ defined in Definition \ref{def:MLG}, a pure strategy for player $P_i \in \mathcal{N}$ is a mapping $q_i: \mathcal{H}_i \mapsto A_i$ such that $q_i(h) \in A(h)$ for every $h \in \mathcal{H}_i$, which specifies what action player $P_i$ will take each time it is his/her turn to play based on the information $h \in \mathcal{H}_i$ he/she possesses. The set of all possible pure strategies for player $P_i$ at MBG $\Gamma$ is then denoted as $Q_i$.
\label{def:local_pure}
\end{definition}

Then, a mixed strategy for player $P_i$ is defined as a probability distribution over the set of his/her pure strategies. 
\begin{definition}[Micro Mixed Strategy]
    Consider the MBG $\Gamma$ defined in Definition \ref{def:MLG}, a mixed strategy for player $P_i \in \mathcal{N}$ is a probability distribution over all of the player $P_i$'s pure strategies, i.e., $\mu_i \in \Delta(Q_i)$.
\label{def:local_mixed}
\end{definition}

\begin{definition}[Micro Behavior Strategy]
    Consider the MBG $\Gamma$ defined in Definition \ref{def:MLG}, a behavior strategy for player $P_i \in \mathcal{N}$ is a mapping $\sigma_i: \mathcal{H}_i \mapsto \Delta(A_i)$, which assigns to each information set $h \in \mathcal{H}_i$ a probability measure on the set $A(h)$. The set of all admissible behavioral strategies of player $P_i$ at MBG $\Gamma$ is denoted as $\Sigma_i$.
\label{def:local_behavior}
\end{definition}
Following Kuhn's theorem in \cite{kuhn1953extensive}, in every MBG $\Gamma$ in extensive form, if player $P_i \in \mathcal{N}$ has ``perfect recall'' as in Assumption \ref{assump:perfect_recall} below, then for every micro mixed strategy there exists an equivalent micro behavior strategy, and vice versa. Hence, we will assume perfect recall and restrict our attention to micro behavior strategies $\sigma_i \in \Sigma_i$, simply called ``strategies'', for the subsequent analysis.

\begin{assumption}[Perfect Recall]
    Each player knows whether he/she chose previously: if $x \in H(x')$, then $x \not\prec x'$. In addition, each player also knows whatever he/she know previously, including his/her previous actions: if $x, x', x'' \in I^{-1}(P_i), x \prec x'$, and $H(x') = H(x'')$, then $H(x)$ includes some predecessor of $x''$ at which the same action was chosen as was chosen at $x$; more formally, $p(x'') \cap H(x) = \{x^0\}$, and if $x=p_n(x')$ and $x^0=p_{n'}(x'')$, then $\alpha(p_{n-1}(x'))=\alpha(p_{n'-1}(x''))$.
\label{assump:perfect_recall}
\end{assumption}

Then, given the nature’s fixed policy $\sigma_c \in \Delta(\mathcal{W})$ (if any) and the strategy profile of the attacker and the defender, i.e., $\Phi=(\sigma_A, \sigma_D, \sigma_c)$, we define $\tau: \mathcal{Z} \mapsto [0, 1]$, a probability measure on the set $\mathcal{Z}$ of game outcomes, as the outcome probability. That is, we use $\tau(z)$ to denote the probability of reaching outcome $z \in \mathcal{Z}$ as 
\begin{equation}
    \tau(z|\Phi)=\sum_{z \in \mathcal{H}^{z}}\sigma_c(p_{l(z)}(z))\left[\Pi_{l=1}^{l(z)}\sigma_{I(p_l(z))}(\alpha(p_{l-1}(z)))\right].
\label{eq:tau_v_z}
\end{equation} 
The expectation operator using $\tau(\cdot|\Phi)$ is denoted as $\E_{\Phi}$. In particular, we use $u_A(\Phi) = u_A(\sigma_A, \sigma_D, \sigma_c) = \E_{\Phi}[r_A(z)]$ to represent player $P_A$'s expected utility from the strategy profile $\Phi=(\sigma_A, \sigma_D, \sigma_c)$. Similarly, $u_D(\Phi) = u_D(\sigma_A, \sigma_D, \sigma_c) = \E_{\Phi}[r_D(z)]$ is player $P_D$'s expected utility.
Building on the notion of base game outcome probability, we now introduce the solution concept for the base game. In game theory, the concept of equilibrium naturally lends itself to the analysis of strategic interactions in steady state within the system. A Nash Equilibrium (NE) in the MBG represents a solution where no player has an incentive to deviate from their chosen strategy. Formally, NE is defined as follows.
\begin{definition}[Nash Equilibrium (NE) in the Micro Base Game]
    For the MBG $\Gamma$ defined in Definition \ref{def:MLG}, given the system randomness $\sigma_c \in \Delta(\mathcal{W})$ over the set of initial states $\mathcal{W}$, a strategy profile $(\sigma^{*}_A, \sigma^{*}_D)$, with $\sigma^{*}_A \in \Sigma_A$ for the player $P_A$ and $\sigma^{*}_D \in \Sigma_D$ for the player $P_D$ is a Nash equilibrium if 
    \begin{equation}
        u_i(\sigma^{*}_i, \sigma^{*}_{-i}, \sigma_c) \geq u_i(\sigma_i, \sigma^{*}_{-i}, \sigma_c)
    \end{equation} for all admissible strategies $\sigma_i \in \Sigma_i$ and for all $P_i \in \mathcal{N}$, where $u_i(\sigma_i, \sigma_{-i}, \sigma_c)$ is the expected utility for player $P_i$ of outcome generated following the strategy profile $\Phi=(\sigma_A, \sigma_D, \sigma_c)$.
    \label{def:ne}
\end{definition}
To solve the game in practice, we consider a refinement of NE tailored for sequential games: the Subgame Perfect Nash Equilibrium (SPNE). It is worth noting that SPNE not only satisfies the conditions of NE but ensures that strategies form an equilibrium in every possible subgame of the overall game. The formal definition is provided below.

\begin{definition}[Subgame Perfection]
    Consider the MBG $\Gamma$ defined in Definition \ref{def:MLG}, a subgame $\Gamma^{'}$ of $\Gamma$ consists of a subset $\mathcal{Y}$ of the nodes $\mathcal{T}$ containing a single non-terminal node $x$ and all of its successors, which has the property that if $y \in \mathcal{Y}, y' \in H(y)$ then $y' \in \mathcal{Y}$, and information sets, feasible moves, and payoffs at terminal nodes as in the MBG $\Gamma$.
\label{def:SP}
\end{definition}

\begin{definition}[Subgame Perfect Nash Equilibrium (SPNE)]
    Consider the MBG $\Gamma$ defined in Definition \ref{def:MLG}, given the system randomness $\sigma_c \in \Delta(\mathcal{W})$ over the set of initial states $\mathcal{W}$, a strategy profile $(\sigma^{*}_A, \sigma^{*}_D)$, with $\sigma^{*}_A \in \Sigma_A$ for the player $P_A$ and $\sigma^{*}_D \in \Sigma_D$ for the player $P_D$ is a subgame perfect Nash equilibrium of $\Gamma$ if it induces a Nash equilibrium in every subgame as defined in Definition \ref{def:SP} of $\Gamma$.
\end{definition} 

For every finite micro-base game $\Gamma$ with fixed system randomness $\sigma_c \in \Sigma_c$, the game admits an SPNE in mixed or behavioral strategies under the assumption of perfect recall, even when players have imperfect information. Moreover, the game with perfect information has an SPNE in pure strategies \cite{maschler2020game}. Since the entire MBG $\Gamma$ is always a subgame of itself, any SPNE for $\Gamma$ must also be an NE defined in Definition \ref{def:ne} for $\Gamma$. 

Hence, SPNE or NE for the MBG $\Gamma$ can then be solved using ``backward induction'' as the game-theory version of the dynamic programming principles, which starts at the end (outcomes/terminal nodes) of the game and then works back to the front.

\section{The Macro-Level Game}

In many real-world scenarios, such as cyber warfare or cyber deception, players may not only participate in a single base game. Instead, they operate across a sequence or set of related games over time. It is often impractical to assume that interactions conclude after the outcome of one base game. For example, an attacker who fails in one attempt may revise their tactics and try again, while a defender may intentionally incur short-term losses to gain intelligence or set traps in order to delay or mislead the adversary for long-term strategic benefit. These types of scenarios reflect the persistent and adaptive nature of adversarial behavior, making it necessary to reason across multiple stages of interactions rather than focusing solely on individual ones.

To support such reasoning and gain higher-level strategic insight, a macro-level game with lower resolution can be constructed by abstracting and synthesizing a set of micro-level base games. This abstraction improves scalability and tractability. As the number of states and detailed interactions increases, solving a set of completely specified micro-level base games simultaneously becomes computationally expensive. In addition, macro-level representations can enhance decision efficiency by focusing on system-level outcomes such as resource allocation, timing of interventions, and overall progress towards the long-term objectives. This broader view enables the prioritization of critical decisions while avoiding unnecessary complexity from fine-grained actions at every step.


\subsection{Game Construction}
Let $\mathcal{S}$ denote a set of micro-level base games considered by players with conflict of interest, where each $\Gamma^s \in \mathcal{S}$ is a micro-base game (MBG) as defined in Definition \ref{def:MLG}, and the superscript $s$ serves to distinguish individual base games. As the superscript $s$ can be used to represent the MBG $\Gamma^s$, we will use $s$ and $\Gamma^s$ interchangeably. However, $s$ will more commonly be used as an abstract representation of the base game $\Gamma^s$ in the macro-level formulation, while $\Gamma^s$ will be reserved for the detailed micro-level formulation. Define $\mathcal{E} \subseteq \mathcal{S} \times \mathcal{S}$ as a set of directed edges representing the relational structure or interdependencies among the base games. The macro-level game topology is then characterized by the directed graph $G = \langle\mathcal{S}, \mathcal{E}\rangle$, where vertices $\mathcal{S}$ correspond to the set of base games and edges $\mathcal{E}$ indicate their relations (feasibility of playing one after the other). With this, the interactive decision-making by players at the macro-level across different MBGs can be modeled as a Markov game.

\begin{definition}[Macro-Strategic Game (MSG)]
\label{def:MSG}
    The Macro Strategic Game (MSG) is defined as a Markov game $M_G$, where the subscript $G$ denotes the directed graph $G = \langle \mathcal{S}, \mathcal{E} \rangle$ that represents the macro-level topology. Each vertex $s \in \mathcal{S}$ corresponds to a micro-base game $\Gamma^s$ as defined in Definition \ref{def:MLG}, and each edge $(s, s') \in \mathcal{E}$ captures the relationship between the base games. The MSG is formally represented by the tuple $M_G=\langle \mathcal{N}, \mathcal{S}, \mathcal{A}^M_A, \mathcal{A}^M_D, T, R_A, R_D, \gamma \rangle$, where each component represents:
    \begin{itemize}
        \item $\mathcal{N}=\{P_A, P_D\}$ represents the set of players, where $P_A$ typically represents the attacker (malicious entity) and $P_D$ denotes the defender.
        \item $\mathcal{S}$ is the state space. Each state $s \in \mathcal{S}$ represents an MBG $\Gamma^s$ defined in Definition \ref{def:MLG}.
        \item $\mathcal{A}^M_i=\mathcal{E}$ represents the action set for player $P_i$. The connections between vertices, corresponding to the directed edges in the graph, form the attacker's action space for exploration, and the defender's action space for cutting or securing those connections.
        \item $T: \mathcal{S} \times \mathcal{A}^M_A \times \mathcal{A}^M_D \mapsto \Delta(\mathcal{S})$ is the transition function controlled by the current state and the joint actions of the players, which captures the probability of transitioning from the base game to the games.
        \item $R_i:\mathcal{S} \times \mathcal{A}^M_A \times \mathcal{A}^M_D \times \mathcal{S} \mapsto \mathbb{R}$ is the immediate payoff function for player $P_i$. 
        \item $\gamma \in [0, 1]$ is the discounting factor. 
    \end{itemize}
\end{definition}
For simplicity, we define the transition probability in this work as follows. For every state $s \in \mathcal{S}, a_A \in \mathcal{A}^M_A$, and $a_D \in \mathcal{A}^M_D$, 
\begin{align*}
    T(s'|s, a_A=(s, v), a_D=(s, v'))=&\begin{cases}
    1, & \text{if } v=s, s'=s, \\
    1, & \text{if } v\neq s, v=v', s'=s, \\
    \lambda_A, & \text{if } v\neq s, v\neq v', s'=v, \\
    1-\lambda_A, & \text{if } v\neq s, v\neq v', s'=s, \\
    0, & \text{if } v\neq s, v=v', s'=v, \\
    0, & \text{otherwise}.
    \end{cases}
\end{align*}
If the attacker chooses to remain at the same vertex, the self-loop edge will lead to the same state with probability one. If the attacker chooses an outgoing edge to move to another vertex and the defender does not secure that edge, the attempt succeeds with probability \(\lambda_A \in [0, 1]\), representing the attacker’s capability. If the attempt fails, the attacker remains at the current vertex. However, if the attacker attempts to use an outgoing edge that the defender is securing, the attacker will remain in the same state with probability one.

Since the attacker's gain is often the defender's loss, we adopt a zero-sum setting. The attacker receives a positive reward upon entering a vertex, with the reward based on the vertex's importance. In contrast, staying at the same vertex, indicating either a failed move or insufficient information from the MBG, results in a negative penalty. Hence, the attacker's utility function can be defined as follows:
\begin{equation*}
    R_A(s, a_A, a_D, s')=\begin{cases}
    \beta, &  \text{if} \ s'=s\\
    \nu(s'), & \forall s' \in \mathcal{S} \setminus \{s\},
    \end{cases}
\end{equation*} where $\beta \in \mathbb{R}_{-}$ is a penalty for the attacker staying at the same vertex without any progress, and $\nu: \mathcal{S} \mapsto \mathbb{R}_{+}$ is the reward for entering another state. With this, the defender's utility is then defined as $R_D(s, a_A, a_D, s')=-R_A(s, a_A, a_D, s')$.

\subsection{Macro-Strategies and Solution Concepts}

Players in the MSG can optimize their strategy against the opponent by leveraging all available information up to the point of decision-making, which is known as the behavioral strategy. In this work, we specifically focus on the Markov (mixed) strategy, a particular type of behavioral strategy, for both players.

\begin{definition}[Macro-Attack Strategy]
\label{def:GAS_A}
    Consider the MBG defined in Definition \ref{def:MLG} and the MSG defined in Definition \ref{def:MSG}. The macro attack strategy in MSG is a mapping from the state space $\mathcal{S}$ to the macro action space $\mathcal{A}^M_A$, i.e., $\pi_A: \mathcal{S} \mapsto \Delta(\mathcal{A}^M_A)$. 
\end{definition} 

\begin{definition}[Macro-Defense Strategy]
\label{def:GAS_D}
    Consider the MBG defined in Definition \ref{def:MLG} and the MSG defined in Definition \ref{def:MSG}. The macro defense strategy in MSG is a mapping from the state space $\mathcal{S}$ to the macro action space $\mathcal{A}^M_D$, i.e., $\pi_D: \mathcal{S} \mapsto \Delta(\mathcal{A}^M_D)$. 
\end{definition} 

At this macro stage, players do not care about detailed interactions (i.e., action sequences) within each MBG $\Gamma^s \in \mathcal{S}$. Instead, they focus on higher-level strategic interactions across games. In a general zero-sum game, the saddle-point equilibrium (SPE) is the most fundamental solution concept. As the attacker’s gain is considered the defender’s loss, we can then use a single payoff function $R(s, a_A, a_D, s') = R_A(s, a_A, a_D, s')=-R_D(s, a_A, a_D, s')$ to construct the game at each decision point. In this case, player $P_A$ aims to maximize the outcome of the game, while $P_D$ aims to minimize. The macro game $M_G$ is played in discrete time over a finite horizon, i.e., $k = 1, 2, \cdots, K$. Starting from the initial state $s_0 \in \mathcal{S}$, player $P_A$ ($P_D$) aims to find the (stationary) strategies $\pi_A$ ($\pi_D$) that maximize (minimize) the expected sum of the discounted payoff:
\begin{align*}
    U(s^0,\pi_A,\pi_D)= \E\left[ \sum_{k=0}^K \gamma^k R^k(s^k, a^{k}_A, a^{k}_D, s') \mid s^0, \pi_A,\pi_D \right],
\end{align*} where $R^k$ is the payoff at stage $k$ and the expectation is taken over the players’ strategy profile $(\pi_A,\pi_D)$.
\begin{definition}[Saddle-Point Equilibrium (SPE) in the Macro Game]
    Consider the MSG $M_G$ defined in Definition \ref{def:MSG}, a saddle-point equilibrium in (stationary) strategies is a strategy pair $(\pi^*_A,\pi^*_D)$ such that, for any stationary strategies  $\pi_A \in \Delta(\mathcal{A}^M_A),\pi_D \in \Delta(\mathcal{A}^M_D)$,
\begin{equation}
\label{eq:spe}
    U(s,\pi_A,\pi^*_D)\leq U(s,\pi^*_A,\pi^*_D)\leq U(s,\pi^*_A,\pi_D),
    \  \forall s\in \mathcal{S}.
\end{equation}
\end{definition} Then, the existence of the saddle-point (stationary) strategy in zero-sum  $\gamma$-discounted stochastic games is established by Shapley \cite{shapley1953stochastic}.

As the payoff of each player will depend on what actions all the players (the attacker and the defender) will take in the future, the value function that estimates the expected reward for the player to be in a given state will depend on the strategies of all the players. In the zero-sum setting, we will focus on the value function of the attacker in the content that follows.
Denote the macro-strategy profile for the attacker and the defender as $\Pi=(\pi_A, \pi_D)$, then for a given state $s \in \mathcal{S}$, the value function under $\Pi$ is given by
\begin{align*}
    V^{\Pi}(s) = &\sum_{a_A \in \mathcal{A}^M_A}\pi_A(a_A|s)\sum_{a_D \in \mathcal{A}^M_D}\pi_D(a_D|s)\\
    &\sum_{s' \in \mathcal{S}}T(s'|s, a_A, a_D)\left[R(s, a_A, a_D, s') + \gamma V^{\Pi}(s')\right]
\end{align*} which is the expected payoff starting from state $s$ and following macro strategy profile $\Pi$. Then, based on the value function, one way to find the optimal strategies for zero-sum Markov games is by formulating
the problem using mathematical programming, as in \cite{filar2012competitive}. 

\section{The Multi-Resolution Game}
From Definition~\ref{def:MLG} (for MBG) and Definition~\ref{def:MSG} (for MSG), we observe the presence of two levels of resolution. In the MSG, players do not consider the detailed interactions within each state $s \in \mathcal{S}$, where $s$ represents a lower-resolution (coarse) abstraction of the underlying base game $\Gamma^s$, but instead focus solely on the value of the state. That is, a penalty for remaining in the current state ($\beta$), or a utility value $\nu(s')$ for entering a new state $s' \in \mathcal{S}$. In contrast, the higher-resolution MBG $\Gamma^s$ captures the full detail of player interactions, where every move in the extensive-form game tree matters.

Hence, there are $2^{|\mathcal{S}|}$ possible resolution configurations for the set of micro-level base games $\mathcal{S}$, where each base game can be treated either at the micro (high-resolution) or macro (low-resolution) level. We define the set of all such resolution configurations as $\mathcal{C}_{\mathcal{S}}$, with each element denoted as $C \in \mathcal{C}_{\mathcal{S}}$. For example, if $|\mathcal{S}|=2$, then $|\mathcal{C}_{\mathcal{S}}|=4$ with $\mathcal{C}_{\mathcal{S}}=\{\{s^1, s^2\}, \{s^1, \Gamma^2\}, \{\Gamma^1, s^2\}, \{\Gamma^1, \Gamma^2\}\}$, where $s^i$ represents low-resolution and $\Gamma^i$ is for high-resolution. With resolution configurations, the remaining challenge lies in how to integrate games across different resolutions together.

\begin{definition}[Resolution Configurations]
\label{def:reso_config}
Let $\mathcal{S}$ be the set of MBGs defined in Definition \ref{def:MLG}, and let the corresponding MSG be defined as in Definition \ref{def:MSG}. The set of resolution configurations, denoted by $\mathcal{C}_{\mathcal{S}}$, consists of all possible combinations of configurations on $\mathcal{S}$, where each configuration specifies whether a base game $\Gamma^s \in \mathcal{S}$ is represented at the micro level $\Gamma^s$ or abstracted at the macro level $s$.
\end{definition}

\begin{remark}
    Note that MSG $M_G$ can be interpreted as the resolution configuration in which all base games in $\mathcal{S}$ are treated at the macro (low-resolution) level. We can refer to it as a ``Completely Abstracted Game'' (CAG), where the resolution is denoted as $C_l \in \mathcal{C}_{\mathcal{S}}$.
\end{remark}

\subsection{Zoom-In Operation}

From the defensive entity's point of view, we can always begin with the CAG and try to ``zoom in'' a specific state $s$ to the high-resolution MBG $\Gamma^s$ to obtain deeper insight into the malicious entity's strategies and sequence of actions. This may require the following consistencies from low to high resolutions.


Recall that $\mathcal{Z}^s$ denotes the set of outcome nodes for the game $\Gamma^s$. In this work, we assume that outcomes for $\Gamma^s$ involve either remaining at the current game $\Gamma^s$ or transitioning to an adjacent, reachable game $s'$ within the graph (i.e., $(s, s') \in \mathcal{E}$). Thus, with a slight abuse of notation, we can denote $\mathcal{Z}^s = \{s' \mid s' \in \mathcal{S}, (s, s') \in \mathcal{E}\}$. 
Then, the utility of each outcome $z \in \mathcal{Z}^s$ now describes the expected reward of moving to the next game (state) $s$ in the MSG. Thus, given the macro strategy $\Pi$, the attacker’s utility functions of reaching outcome $z \in \mathcal{Z}^s$ in the MBG $\Gamma^s$ are updated as
\begin{equation}
    \begin{aligned}
    r^s_A(z=s') = \sum_{a_D \in \mathcal{A}^M_D}\pi_D(a_D|s)
    \sum_{s'' \in \mathcal{S}}\Big[&T(s''|s, a_A=(s, s'), a_D)\\ &\left[R(s, a_A, a_D, s'') + \gamma V^{\Pi}(s'')\right]\Big].
\end{aligned}
\label{eq:utility_consist}
\end{equation}
Similarly, the utility of the defender is the opposite of that of the attacker. That is, $r^s_D(z=s')=-r^s_A(z=s')$. Then, we can now define the ``zoom in'' operator for the multi-resolution game.

\begin{definition}[Zoom-In Operation]
Consider the set $\mathcal{S}$ of MBGs as defined in Definition \ref{def:MLG}, the corresponding MSG defined in Definition \ref{def:MSG}, and the set of resolution configurations $\mathcal{C}_{\mathcal{S}}$ defined in Definition \ref{def:reso_config}. The ``zoom-in operator'', denoted $\texttt{OP}_{\texttt{in}}$, is a mapping that takes a current resolution configuration with a specific game $s \in \mathcal{S}$, and returns a new configuration in which the resolution of $s$ is set to the micro level. Formally,
$\texttt{OP}_{\texttt{in}} : \mathcal{C}_{\mathcal{S}} \times \mathcal{S} \to \mathcal{C}_{\mathcal{S}}$, with the outcome utilities updated by \eqref{eq:utility_consist} given the macro-strategy profile $\Pi$.
\label{def:zoom-in}
\end{definition}

\subsection{Zoom-Out Operation}

Following the ``zoom-in'' operation, a corresponding ``zoom-out'' operation naturally arises. If we aim to ``zoom out'' the MBG $\Gamma^s$ to low-resolution $s$, given the micro strategy profile $\Phi^s=(\sigma^s_A, \sigma^s_D, \sigma^s_c)$, the macro attack strategy $\pi_A(\cdot|s)$ needs to be updated in order to be ``consistent'' with the outcome probability in MBG, which is given by
\begin{equation}
    \pi_A(a_A|s)=\pi(a_A=(s, z)|s)=\tau^s(z|\Phi^s),\ \ \forall z \in \mathcal{Z}^s,
\label{eq:outcome_consist}
\end{equation} where $\tau^s(z|\Phi^s)$ is the outcome probability in   \eqref{eq:tau_v_z}.
As a result, the ``zoom out'' operator for the multi-resolution game can be defined as follows.

\begin{definition}[Zoom-Out Operation]
Consider the set $\mathcal{S}$ of MBGs as defined in Definition \ref{def:MLG}, the corresponding MSG defined in Definition \ref{def:MSG}, and the set of resolution configurations $\mathcal{C}_{\mathcal{S}}$ defined in Definition \ref{def:reso_config}. The ``zoom-out operator'', denoted $\texttt{OP}_{\texttt{out}}$, is a mapping that takes a current resolution configuration with a specific game $\Gamma^s \in \mathcal{S}$, and returns a new configuration in which the resolution of $\Gamma^s$ is set to the macro level. Formally,
$\texttt{OP}_{\texttt{out}} : \mathcal{C}_{\mathcal{S}} \times \mathcal{S} \to \mathcal{C}_{\mathcal{S}}$, with the macro-strategy updated by \eqref{eq:outcome_consist} given the micro-strategy profile $\Phi^s=(\sigma^s_A, \sigma^s_D, \sigma^s_c)$.
\label{def:zoom-out}
\end{definition}

\subsection{Multi-Resolution for Purple Teaming}

Given the definitions \ref{def:zoom-in} and \ref{def:zoom-out}, a sequence of zoom-in and zoom-out operations can be applied to systematically adjust the resolution of the game. This enables deeper insight into player interactions and facilitates strategic refinement across different levels of granularity. From the defender’s perspective, such operations support purple teaming reasoning. By zooming in, the defender can analyze the attacker’s strategy at the higher resolution and identify equilibrium behaviors within individual micro-base games. Subsequently, the defender can zoom out to re-evaluate and adjust the macro-level strategy, with the goal of achieving an overall strategic advantage even if certain micro-level engagements are lost.

\begin{figure}
    \centering \vspace{-1mm}
    \includegraphics[width=4.8in]{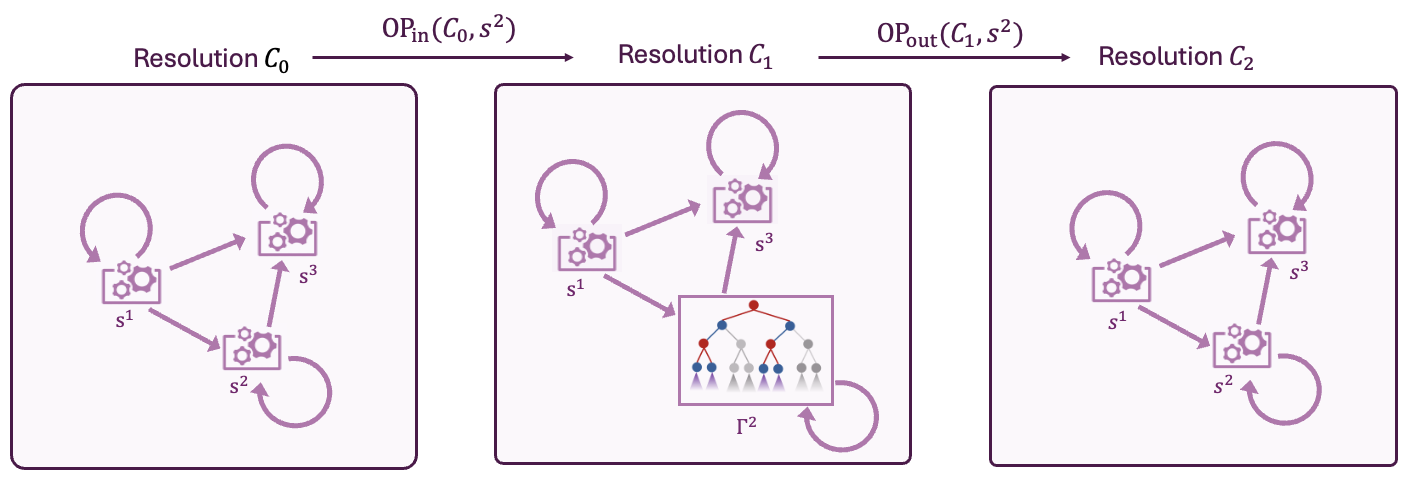}\vspace{-3mm}
    \caption{An illustrative example for multi-resolution operations.}\vspace{-5mm}
    \label{fig:reso_ex}
\end{figure}

\begin{algorithm}
  \caption{Multi-Resolution Operations} \label{algo:exe}
  \begin{algorithmic}[1]
    \State\textbf{Input} MBG set $\mathcal{S}$, initial resolution $C_0$, sequence of resolution operations $\Omega$
    \State $\qquad \quad$ Value function $V^\Pi$ under current macro strategy $\Pi = (\pi_A, \pi_D)$
    \State\textbf{Initialize} $C \leftarrow C_0$
    \For{operation $(\texttt{OP}, s) \in \Omega$}
    \If{$\texttt{OP} = \texttt{OP}_{\texttt{in}}$}
        \State\textbf{Let} the outcome utility for each $z \in \mathcal{Z}^s$ be defined as \eqref{eq:utility_consist}
        \State\textbf{Compute} the micro strategy profile $\Phi^s=(\sigma^s_A, \sigma^s_D, \sigma^s_c)$ for $\Gamma^s$
        \State\textbf{Update} $C \leftarrow \texttt{OP}_{\texttt{in}}(C, s)$
    \ElsIf{$\texttt{OP} = \texttt{OP}_{\texttt{out}}$}
        \State\textbf{Retrieve} micro strategy profile $\Phi^s=(\sigma^s_A, \sigma^s_D, \sigma^s_c)$ for $\Gamma^s$
        \State\textbf{Compute} the macro strategy for each $z \in \mathcal{Z}^s$ according to \eqref{eq:outcome_consist}, update $\Pi$
        \State\textbf{Update} $C \leftarrow \texttt{OP}_{\texttt{out}}(C, s)$
    \EndIf
    \EndFor
    \State \Return configuration $C$ and macro strategy profile $\Pi$
  \end{algorithmic}
  \label{alg:operation}
\end{algorithm}

For instance, consider a set of MBGs of size $|\mathcal{S}| = 3$, and let the initial resolution configuration be $C_0 = C_l = \{s^1, s^2, s^3\}$, where each state $s^i \in \mathcal{S}$ is initially represented at the macro level (low-resolution), forming a completely abstracted configuration (CAG). Then, given a sequence of resolution operations $\Omega = \{(\texttt{OP}_{\texttt{in}}, s^2), (\texttt{OP}_{\texttt{out}}, s^2), (\texttt{OP}_{\texttt{in}}, s^3), (\texttt{OP}_{\texttt{out}}, s^3),
(\texttt{OP}_{\texttt{in}}, s^1)\}$, we can apply them iteratively as follows:
\[
\begin{aligned}
    C_1 &= \texttt{OP}_{\texttt{in}}(C_0, s^2), \quad
    C_2 = \texttt{OP}_{\texttt{out}}(C_1, s^2), \quad
    C_3 = \texttt{OP}_{\texttt{in}}(C_2, s^3), \\
    C_4 &= \texttt{OP}_{\texttt{out}}(C_3, s^3), \quad
    C_5 = \texttt{OP}_{\texttt{in}}(C_4, s^1).
\end{aligned}
\]
where each operation adjusts the resolution of the corresponding state $s^i \in \mathcal{S}$ to either the micro level (via $\texttt{OP}_{\texttt{in}}$) or the macro level (via $\texttt{OP}_{\texttt{out}}$). Note that the zoom-out operation for a specific game can only be applied after a zoom-in operation has been performed on that game. In this case, as illustrated in Figure \ref{fig:reso_ex}, $C_1=\{s^1, \Gamma^2, s^3\}$, where the outcome utilities of $\Gamma^2$ are as in \eqref{eq:utility_consist} based on current macro-strategy profile $\Pi$ and value function $V^{\Pi}$. The micro-strategy profile $\Phi^s=(\sigma^s_A, \sigma^s_D, \sigma^s_c), s=2$ for $\Gamma^2$ is computed. Then, $C_2=\{s^1, s^2, s^3\}$, where the macro-strategy of $\pi_A(\cdot|s^2)$ are updated according to \eqref{eq:outcome_consist} based on $\Phi^2$. The procedure continues similarly for $C_3, C_4$, and $C_5$, and is summarized more generally in Algorithm \ref{alg:operation}. These operations facilitate controlled exploration of strategy refinement and game dynamics at different levels of resolution.
To this end, we can define the multi-resolution operation plan as follows.
\begin{definition}[Multi-Resolution Operation Plan]
Consider the set $\mathcal{S}$ of MBGs as defined in Definition \ref{def:MLG}, the corresponding MSG from Definition \ref{def:MSG}, and the set of resolution configurations $\mathcal{C}_{\mathcal{S}}$ defined in Definition \ref{def:reso_config}. A multi-resolution operation plan, denoted by $\Omega$, is defined as a finite sequence of zoom-in (described in Definitions~\ref{def:zoom-in}) and zoom-out (described in Definitions~\ref{def:zoom-out}) operations. Each zoom-out operation in the sequence needs to be preceded by a corresponding zoom-in operation on the same game.
\label{def:multi-op}
\end{definition}

\begin{figure}
    \centering \vspace{-1mm}
    \includegraphics[width=4.5in]{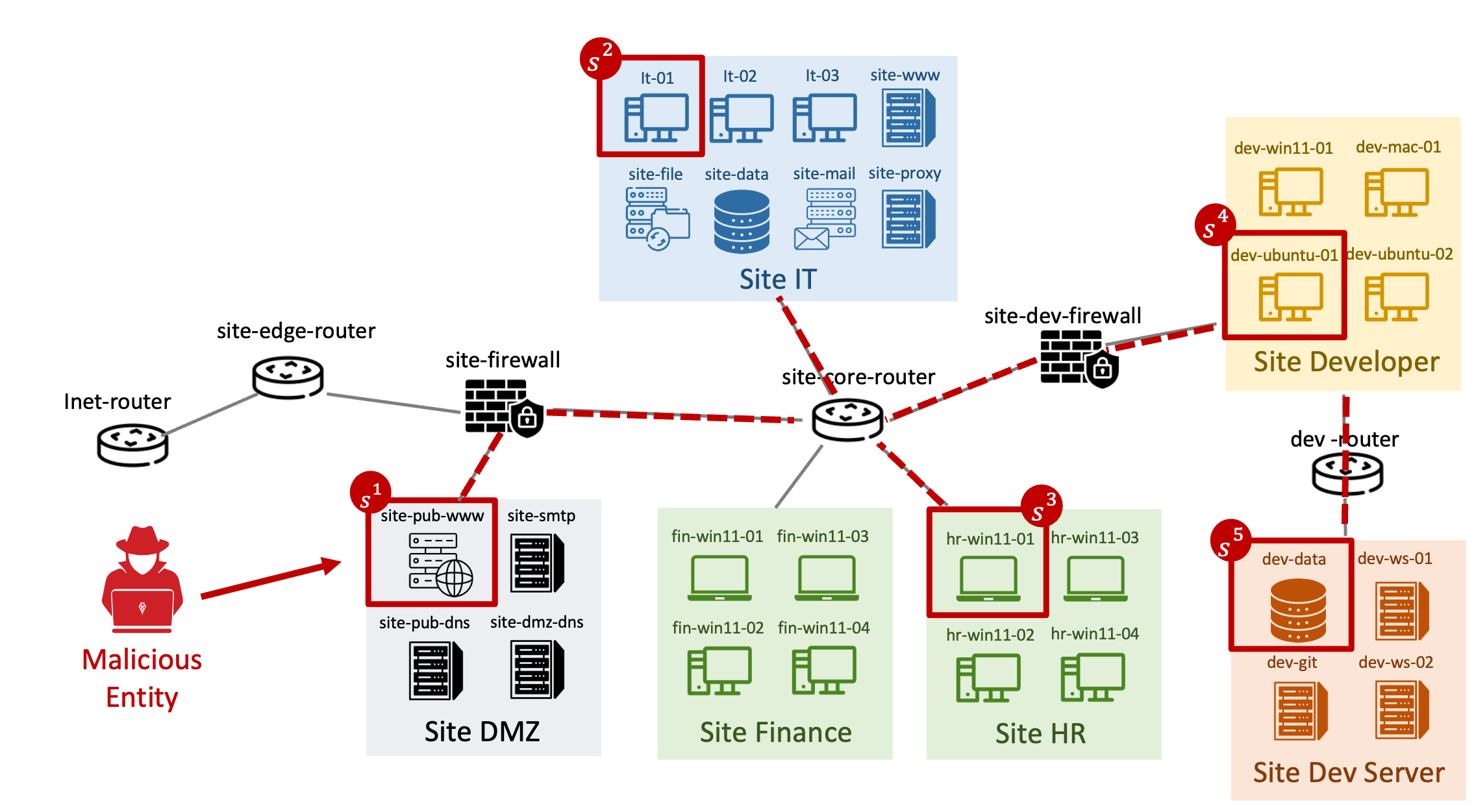}
    \caption{An enterprise network for case study. A possible attack path contains five nodes, including the web server in the DMZ site, devices in the IT site, user devices in the human resource or finance site, the devices in the developer site, and the critical asset located on the developer server.}\vspace{-1mm}
    \label{fig:net}
\end{figure}

The multi-resolution operation plan can be tailored to the needs of a specific system. While it may not always lead to performance improvement, especially when the existing macro-level strategy profile is already good enough. Instead, it provides a valuable mechanism for selectively increasing model fidelity. What such a plan offers is the ability to zoom in on specific parts of the system to examine whether local interactions, when modeled in greater detail, reveal strategic nuances that may otherwise be overlooked. This includes identifying latent vulnerabilities, refining subgame-level strategies, or uncovering local inefficiencies that could inform more robust macro-level decision-making.

\section{Case Study}

\begin{figure}
    \centering \vspace{-1mm}
    \includegraphics[width=3.5in]{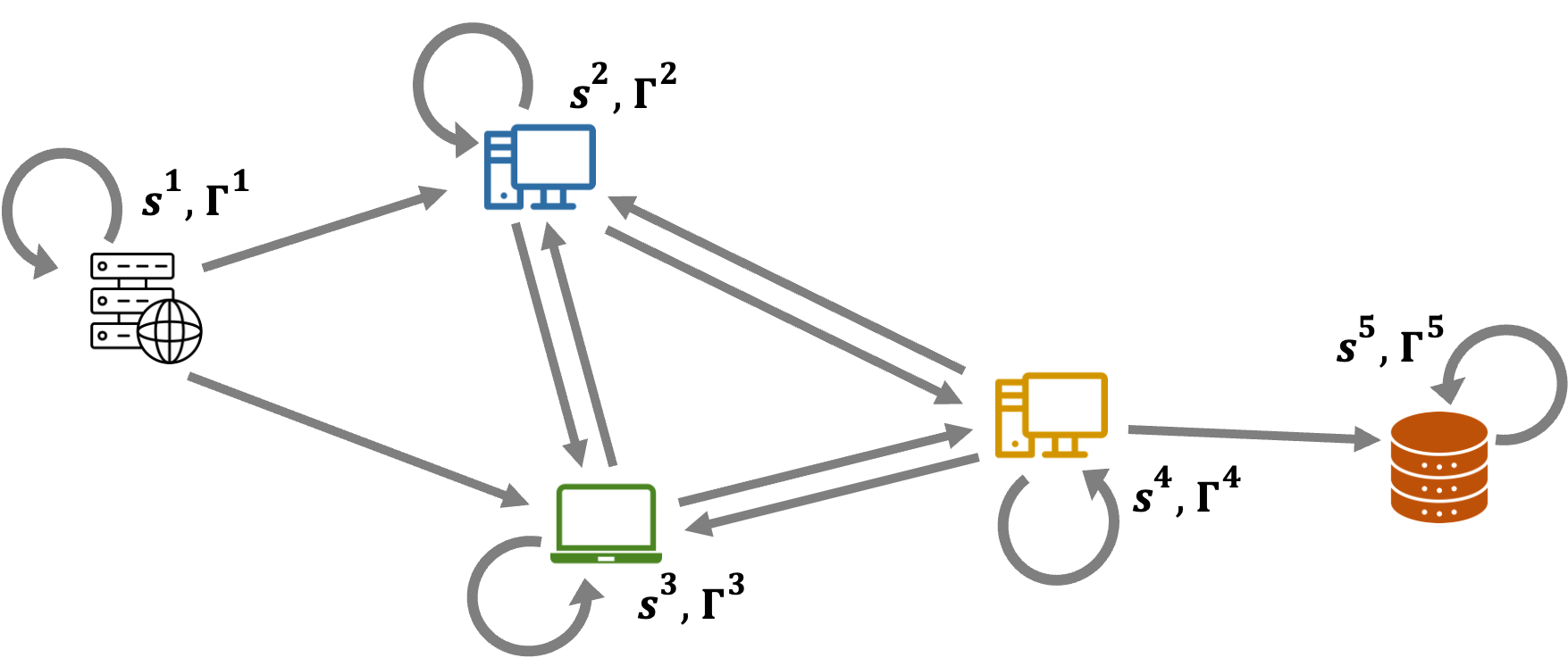}
        \caption{The MSG representation constructed from the enterprise network topology in Figure~\ref{fig:net}. Each vertex corresponds to a MBG associated with a distinct subsystem in the network and can be modeled at two levels of resolution: low-resolution (denoted \(s_i\)) for strategic abstraction and high-resolution (denoted \(\Gamma_i\)) for detailed tactical modeling. 
        }
   \vspace{-5mm}
    \label{fig:msg}
\end{figure}

\begin{figure}
    \centering \vspace{-4mm}
    \includegraphics[width=4.8in]{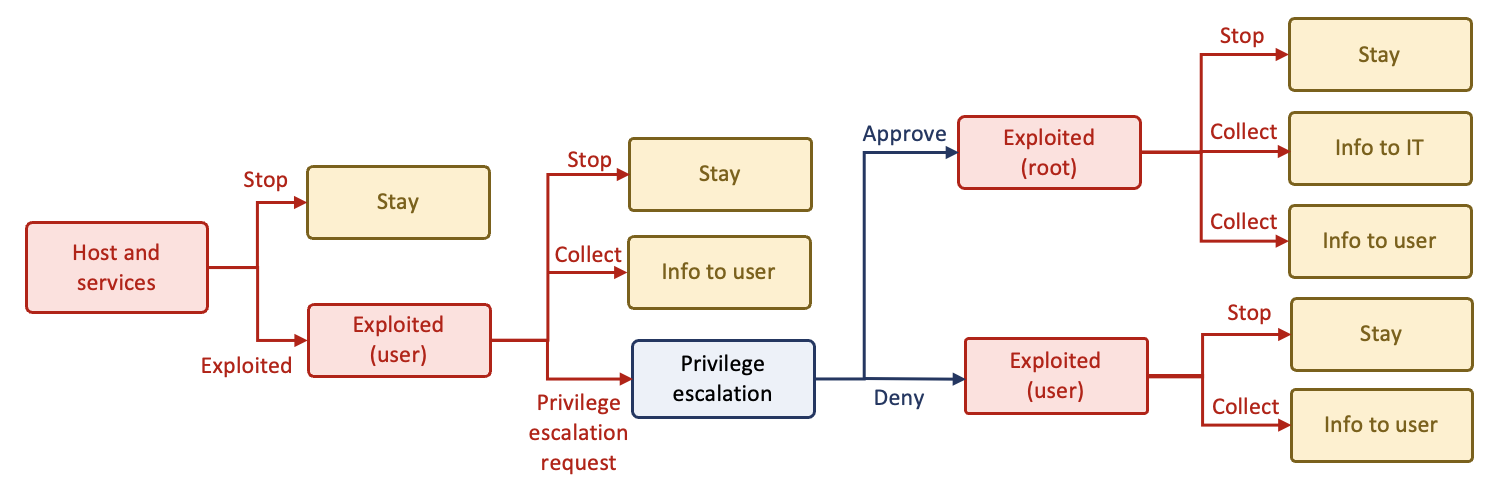}
        \caption{Extensive-form game tree for the MBG \(\Gamma_1\) associated with vertex \(s_1\), representing the web server in the DMZ site. Red and blue nodes indicate attacker and defender decision points, respectively, while yellow nodes denote terminal outcomes \(z \in Z_1\). This structure captures step-by-step interactions at the tactical layer and encodes possible attacker-defender sequences aligned with realistic techniques and procedures, as informed by the MITRE ATT\&CK framework.}
\vspace{-1mm}
    \label{fig:web_dmz}
\end{figure}

We use an enterprise network topology shown in Figure \ref{fig:net} as an illustrative case study to illustrate the operation of the proposed multi-resolution game \footnote{This work uses the network topology provided by the GAMBiT project’s experimental dataset: \url{https://osf.io/ukdwm/}.}. In this example, one possible attack path contains five nodes, including the web server in the DMZ site, devices in the IT site, user devices in the human resource or finance site, the devices in the developer site, and the critical asset located on the developer server. Note that while the developer's server contains valuable data, decoy files are also deployed on the same server to mislead and deceive malicious entities. The network can then be abstracted and a corresponding MSG with $|\mathcal{S}|=5$ can be constructed as illustrated in Figure \ref{fig:msg}, with the corresponding reward function parameters provided in Table \ref{tab:ex1}. One of the extensive-form game trees of the MBG $\Gamma^s$ corresponding to each vertex $s$ is depicted in Figure \ref{fig:web_dmz}. These game trees are intended to align with representative attack scenarios in the MITRE ATT\&CK framework \cite{strom2018mitre} and can be modified to reflect the structure of specific systems.

\subsection{Baseline Scenarios}
\begin{figure}[h]
    \centering 
    \includegraphics[width=2.6in]{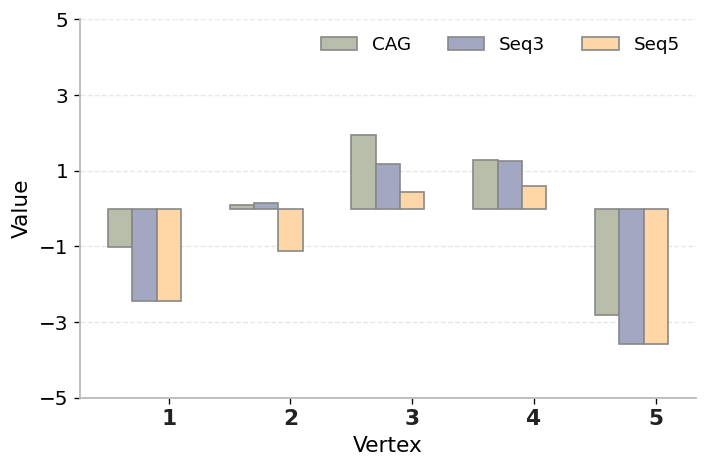}
    \caption{The resulting game (vertex) value for each state $s^i \in \mathcal{S}$ of the multi-resolution game from the MSG in Figure \ref{fig:msg}. CAG refers to the resolution configuration of a completely abstracted game, and Seq3 as well as Seq5 result from sequential zoom-in zoom-out operations. (Here, $\lambda_A=0.6$.)}
    \label{fig:CAG_seq5}
\end{figure}

We consider the following baseline scenarios for comparative purposes:
\begin{itemize}
    \item CAG: In this case, we consider the case of a completely abstracted configuration, where no zoom-in or zoom-out operation is performed.
    \item Seq3: In this case, a sequence of zoom-in zoom-out operations of $s^1$, $s^3$, $s^5$ is performed. That is, $\Omega_{\text{Seq3}}=\{(\texttt{OP}_{\texttt{in}}, s^1), (\texttt{OP}_{\texttt{out}}, s^1), (\texttt{OP}_{\texttt{in}}, s^3), (\texttt{OP}_{\texttt{out}}, s^3),\\ (\texttt{OP}_{\texttt{in}}, s^5), (\texttt{OP}_{\texttt{out}}, s^5)\}$.
    \item Seq5: In this case, a sequence of zoom-in zoom-out operations from $s^1$ to $s^5$ is performed. That is, $\Omega_{\text{Seq5}}=\{(\texttt{OP}_{\texttt{in}}, s^1), (\texttt{OP}_{\texttt{out}}, s^1), (\texttt{OP}_{\texttt{in}}, s^2), (\texttt{OP}_{\texttt{out}}, s^2),\\ \cdots, (\texttt{OP}_{\texttt{in}}, s^5), (\texttt{OP}_{\texttt{out}}, s^5)\}$.
\end{itemize}
The resulting game value for each state $s \in \mathcal{S}$ is shown in Figure \ref{fig:CAG_seq5}. Recall that in this zero-sum setting, the attacker’s gain corresponds directly to the defender’s loss. Therefore, from the defender’s perspective, a lower game value is more favorable, as it indicates reduced utility for the attacker. As illustrated in the figure, the defender is able to achieve a more favorable outcome by sequentially zooming in on certain micro-level base games. This enables the defender to observe and model detailed interactions more precisely, leading to the refinement of (macro) defensive strategies. By incorporating these multi-resolution operations, the defender can more effectively limit the attacker’s success, thus lowering the overall game value associated with the attacker’s strategy.

\begin{table*}[h!]
\label{tab:ex1}
\centering
\caption{Parameters for the reward function (in MSG): At state \(s^5\), the attacker receives a reward of 15 for accessing valuable data and \(-1\) if deceived by decoy data.}
 \begin{tabular}{|c c c c c c c|} 
 \hline 
  &  $\quad\nu(s^1)\quad$ & $\quad\nu(s^2)\quad$ & $\quad\nu(s^3)\quad$ & $\quad\nu(s^4)\quad$ & $\quad\nu(s^5)\quad$ & $\quad\beta\quad$ \\ [0.5ex] 
 \hline \hline 
 value &  1 & 5 & 1 & 10 & 10 & -2\\ 
 \hline
 \end{tabular} \\ [1.0ex]

\end{table*}

\begin{figure}[htp]
\centering \vspace{-1mm}
\subfloat[ \label{fig:seq3_07}]{%
  \includegraphics[width=0.5\textwidth]{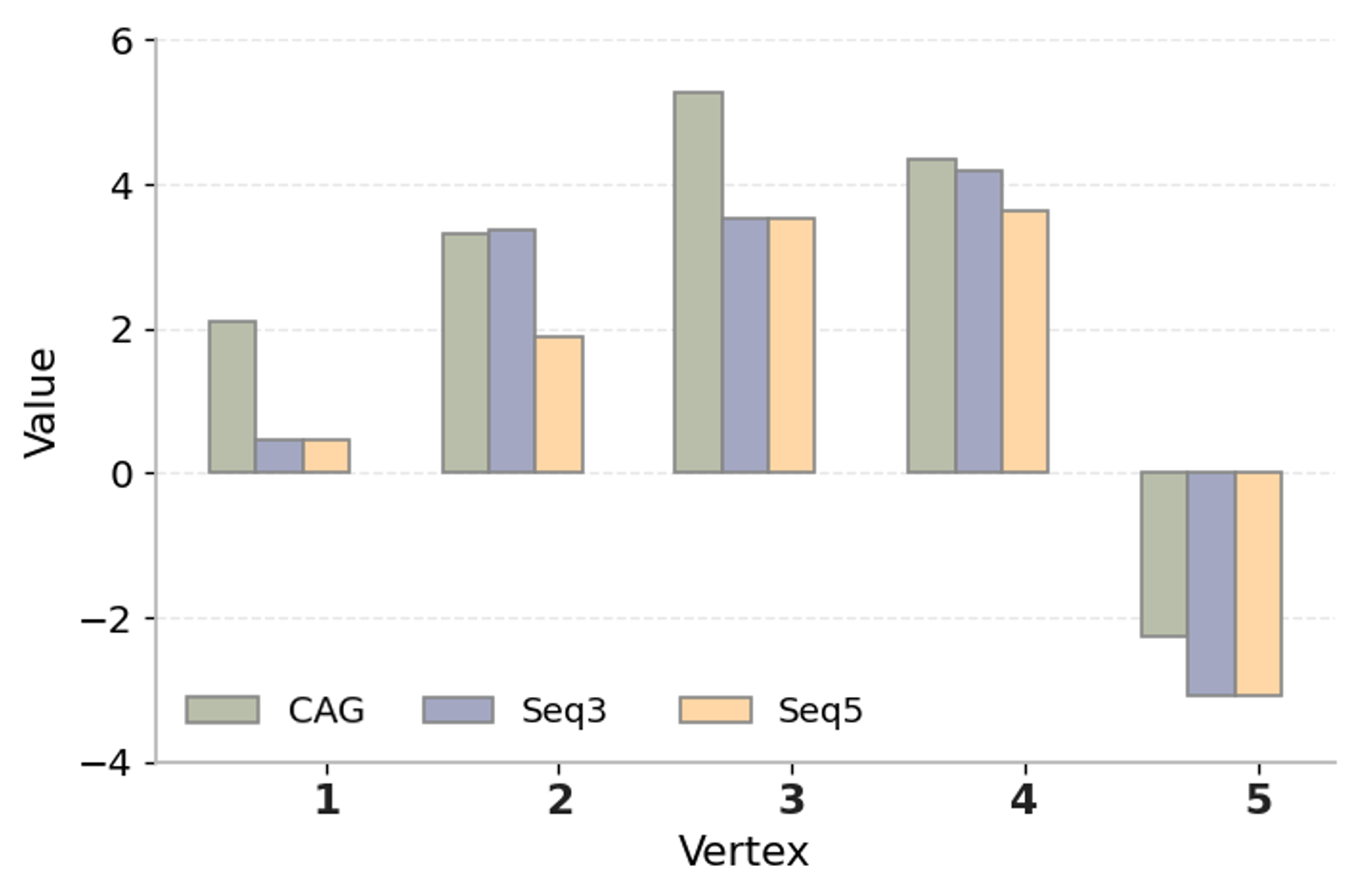}%
}\hfil
\subfloat[ \label{fig:seq3_05}]{%
  \includegraphics[width=0.5\textwidth]{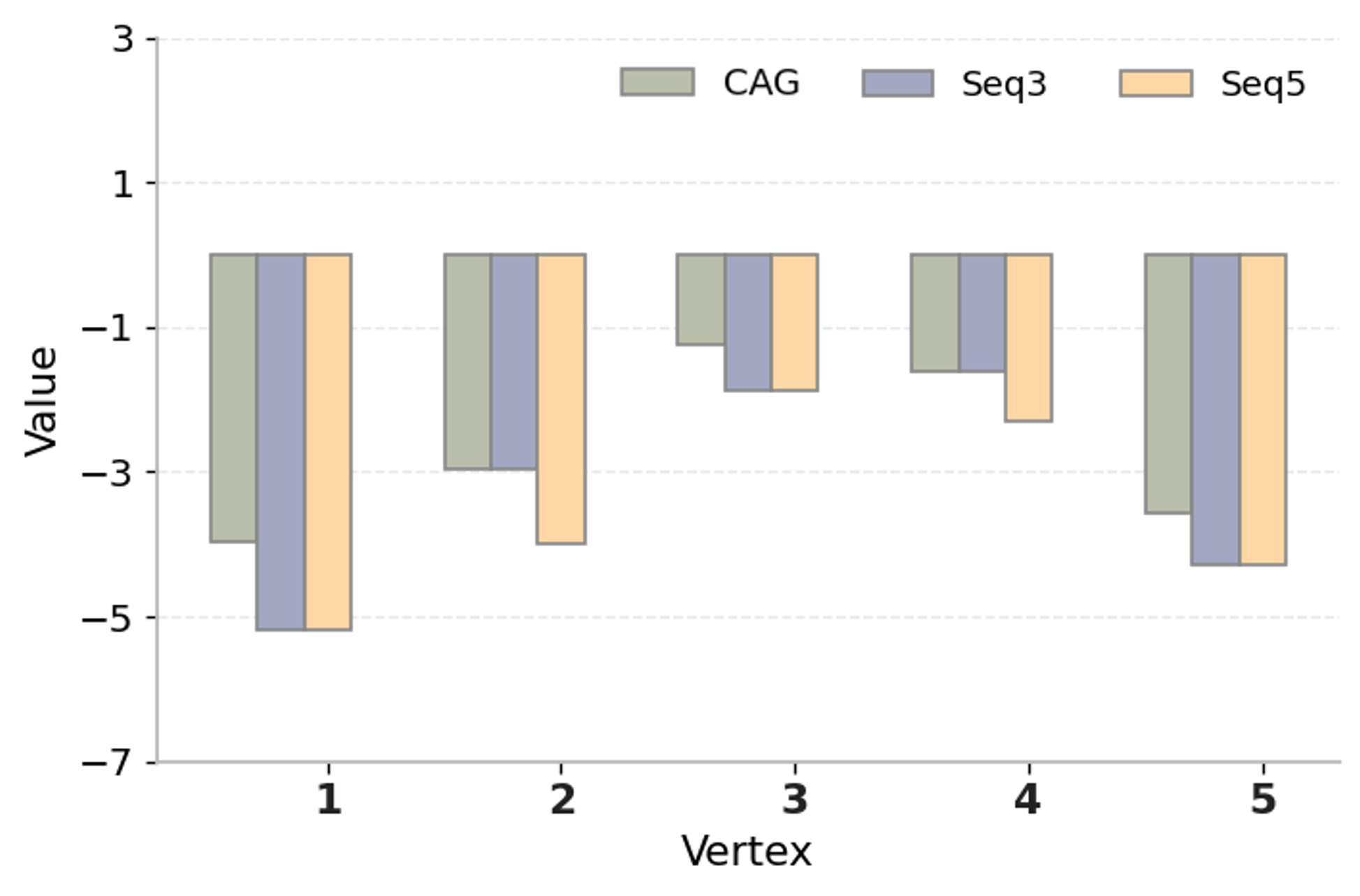}%
}
\caption{The resulting game (vertex) value for each state $s^i \in \mathcal{S}$ of the multi-resolution game is shown for different attacker's capabilities $\lambda_A$: (a) corresponds to $\lambda_A=0.7$, and (b) corresponds to $\lambda_A=0.5$.}
\vspace{-1mm}
\label{fig:seq3}
\end{figure}

\subsection{Different Attacker's Capabilities}
It is worth noting that the transition probability of the MSG also depends on the attacker's capability $\lambda_A$. Hence, we consider $\lambda_A=0.7$ and $0.5$ for the following cases shown in Figure \ref{fig:seq3}.

The results are consistent with intuition, as higher attacker capability generally results in greater value for the attacker. When the attacker’s capability is relatively low, the outcomes from the CAG may already be satisfactory for the defender, so applying sequences of zoom-in and zoom-out operations might be unnecessary unless the defender is specifically interested in the detailed tactics and interactions of a particular MBG. On the other hand, when the attacker is more capable, a better understanding of the interactions within micro-base games at the tactical layer can support more effective overall defensive strategies. This is evident in the comparison between Seq5 and Seq3, where Seq5 provides more favorable outcomes for the defender.

\section{Discussion and Conclusions}
This work has presented a multi-resolution dynamic game framework for enabling cross-echelon decision-making in cyber warfare by systematically integrating tactical and strategic reasoning. At the tactical layer, adversarial interactions are modeled using Micro Base Games (MBGs), formulated as extensive-form game trees that capture fine-grained dynamics such as specific tactics, techniques, and procedures (TTPs). At the strategic layer, a Markov-based Macro Strategic Game (MSG) abstracts these MBGs into lower-resolution states, facilitating scalable reasoning, long-term mission planning, and coordination across interdependent components of the system.

A central contribution of this framework is the introduction of formal zoom-in and zoom-out operations, which enable dynamic adjustment of modeling resolution based on operational needs. These operations support adaptive strategy refinement, cross-layer integration, and purple teaming by allowing selective deep dives into specific MBGs while maintaining global situational awareness. Our case study demonstrates that incorporating high-resolution insights into the macro-level planning process can significantly improve the defender’s strategic outcomes. The framework also addresses the challenge of modeling complexity by leveraging system structure and network topology to mitigate the curse of dimensionality. By decomposing the decision space into resolution-aware layers, it provides a tractable approach for analyzing large-scale adversarial systems.

Looking forward, several research directions emerge. The theoretical foundations of multi-resolution games warrant further development, particularly in understanding equilibrium properties, designing efficient learning algorithms, and characterizing the underlying information structures—especially under partial observability. Moreover, the application of epistemic constraints to guide or limit zoom-in and zoom-out operations opens a path toward cognitively-aware modeling. Beyond cyber warfare, this framework offers a versatile paradigm applicable to other complex domains such as multi-domain operations involving agents across air, land, sea, space, and cyber domains, as well as the analysis of interdependent critical infrastructures. It provides a principled foundation for reasoning across scales in systems-of-systems where local interactions lead to macro-level consequences.


\bibliographystyle{splncs04}
\bibliography{mybibliography}

\end{document}